# Evidence for a quantum phase transition in $Pr_{2-x}Ce_xCuO_{4-\delta}$ from transport measurements

Y. Dagan, M. M. Qazilbash, C. P. Hill, V. N. Kulkarni, and R. L. Greene

*Center for Superconductivity Research, Department of Physics, University of Maryland, College Park, Maryland 20742, USA*

Abstract

The doping and temperature dependence of the Hall coefficient, $R_H$, and *ab*-plane resistivity in the normal state down to 350mK is reported for oriented films of the electron-doped high-$T_c$ superconductor $Pr_{2-x}Ce_xCuO_{4+\delta}$. The doping dependence of $\beta$ ($\rho=\rho_0+AT^\beta$) and $R_H$ (at 350 mK) suggest a quantum phase transition at a critical doping near x=0.165.

The mechanism for the superconductivity (SC) in the high-$T_c$ copper oxides is still a mystery in spite of 17 years of intense research on these materials [1]. Moreover, many of the normal state properties are not understood either. Of the many ideas put forth to explain the cuprates, the existence of a quantum phase transition (QPT) between two phases near a critical doping, $x_c$, and its relationship to the unconventional SC is emerging as an important issue [2, 3, 4]. The nature of these phases and their relation to the origin of the high-Tc SC is unclear at present. Although a QPT occurs at T=0 it is characterized by a "funnel-shaped" region in the p-T phase diagram, where p is the control parameter. [5] The physical properties in this region are governed by quantum fluctuations. A study of some of these physical properties in the electron-doped cuprates, with the doping level being the control parameter, is the focus of the research presented in this paper.

So far, the evidence for a QPT in the hole-doped (p-type) cuprates has been somewhat indirect [6] because the upper critical fields ($H_{c2}$) are extremely large and it is not possible to destroy the SC to examine the normal state properties down to very low temperatures. A recent exception [7] will be discussed later. Because of the lower $H_{c2}$ (10T maximum at optimal-doping), the electron-doped cuprates ($R_{2-x}Ce_xCuO_{4+\delta}$ (RCCO) with R=Nd, Pr, La, Sm) offer a distinct advantage for investigating the consequences of a doping dependent QPT in the cuprates. In the electron-doped (n-type) cuprates an antiferromagnetic (AFM) phase starts at x=0 and persists up to, or into, the SC dome [8]. Recent neutron scattering experiments [9] show an AFM phase for $H>H_{c2}$ in optimally-doped n-type cuprates, but no such phase on the overdoped side. These experiments suggest that AFM is the competing phase with SC in the n-type cuprates. Tunneling measurements give evidence for a normal state pseudogap for dopings within the SC dome [10] which goes away near x=0.17 [11]. It is not known if this pseudogap is similar to that found in the p-type cuprates or if it is related to the AFM order. ARPES experiments on NCCO show that the Fermi surface (FS) changes dramatically as a function of doping, going from a small electron pocket at ($\pi$,0) in underdoped (x=0.04) to a large hole-like FS at optimal doping (x=0.15) [12]. Neither the magnetic measurements nor ARPES have had the energy or doping precision to determine $x_c$. The detailed low temperature transport experiments presented here show that a quantum critical point (QCP) occurs at $x_c$=0.165 ($\pm$0.005) in PCCO.

The normal state transport properties of the cuprates for T>Tc do not follow the behavior expected for conventional metals (Fermi liquid (FL)): the Hall coefficient has a strong T dependence, the *ab*-plane resistivity varies as $\rho \propto T$ (for p-type), $\rho \propto T^2$ (for n-type) up to temperatures greater than 250K. A marginal FL model has been successful in explaining these properties in the hole-doped cuprates (as well as some other properties) [13]. This model is compatible with the notion of a QPT near optimal-doping [3]. In various QCP models there are definite predictions for the behavior of the resistivity and the Hall effect near the QCP [13, 14]. For example, in a *d*-density wave picture a sharp kink in $R_H$ as a function of doping has been predicted [15]. Evidence for some of these predictions has been found in other correlated systems such as heavy fermions [16]. In the cuprates, other suggestive evidence for a QPT under the SC dome has come from the observation of a low T normal state "insulator" to metal crossover as a function of doping in both n-and p-type materials [17, 18].

In this letter we present a comprehensive study of the *ab*-plane resistivity and Hall effect in films of electron-doped PCCO, for many dopings near and within the SC dome. We study the normal state (H>Hc2) from 350mK to 300K with an emphasis on the temperature and doping dependence of these transport properties below 20K. Our results give compelling evidence for a QPT at a critical doping of x=0.165±0.005 (i.e. in the overdoped region of the phase diagram).

PCCO *c*-axis oriented films of various Ce doping concentrations were deposited from stoichiometric targets on (100) oriented $SrTiO_3$ substrates using the pulsed laser deposition (PLD) technique with conditions similar to those reported by Maiser *et al.* [19]. Rutherford back scattering measurements were used to determine the thickness of the films. The minimum-channeling yield obtained was 10%-20% indicating a good epitaxial growth. The low residual resistivity and sharp $T_c$ (see Figure 1) indicate that the films are of better quality than the best previously reported PLD films [18] and comparable to MBE grown films [20]. Since the oxygen content has an influence on both the SC and normal state properties of the material [21] we took extra care in optimizing the annealing process for each Ce concentration. We found that the optimal annealing time, $t_A$, after which $T_c$ stopped changing, increases with increasing Ce doping. We also found that roughly $t_A \propto d^2$, with d being the films' thickness, as expected for a diffusion process. For x≤0.15 we had to simply maximize $T_c$ and avoid decomposition spots detectable in an optical microscope.

Resistivity and Hall voltage were measured up to a magnetic field of 14T and down to a temperature of 0.35K with the field applied perpendicular to the $CuO_2$ planes. At low temperatures the Hall coefficient was measured by taking field scans from –14T to 14T, a field at which the PCCO is well in the normal state even at 0.35K.

In figure 1 we show the *ab*-plane resistivity *versus* temperature at 10T (H>$H_{c2}$) for 0.13≤x≤0.19. First, we note the decrease in the resistivity as the Ce concentration is increased. Another feature appearing in x≤0.15 films is a sign change in dρ/dT (an upturn). The temperature at which the upturn appears decreases with increasing doping. The insert shows the resistive SC transition where $T_c$ has the expected doping dependence [19]. All the films have sharp transitions. The transition width, $\Delta T_c$, measured as the width at half maximum of the peak in dρ/dT, is: $\Delta T_c$=0.3-0.6 K in optimum and overdoped PCCO (increasing with increasing Ce doping), and $\Delta T_c$=2.2 K for x=0.13, all much sharper than previously reported PLD films.

Figure 2a shows typical Hall resistivity $\rho_{xy}$ at various temperatures for the x=0.17 sample. Note that $\rho_{xy}$ is linear as a function of magnetic field above the upper critical field ($Hc_2 \approx$5T for x=0.17). The normal state resistance recovers well below 10T, hence, the measurements taken at high fields are normal state measurements. We calculate the Hall coefficient, $R_H$, from the slope of a linear fit to the data at high fields ($\rho_{xy}=R_H B$). In Figure 2b we show $R_H$ as a function of temperature for various doping levels. Note the strong temperature dependence in the intermediate doping levels even at T<10K. The sign of $R_H$, which in a simple metal corresponds to the type of charge carrier, also changes with doping and temperature. The temperature dependence of $R_H$ (along with other transport properties) was previously interpreted as evidence for two types of carriers for Ce concentrations near optimal (x=0.15) [22]. Although it is not possible to determine the exact oxygen content in the films, $R_H$ at low temperatures depends systematically on Ce doping, which suggests using the low temperatures $R_H$ as a criterion to determine the number of carriers (coming from both Ce and oxygen).

In Figure 3 we show $R_H$ at 0.35K as a function of cerium doping for x=0.11 to x=0.19. An abrupt change in $R_H$ is seen above x=0.16. This is one of the important new results of our work. Previous studies did not measure enough samples to reveal this striking change in $R_H$. $R_H$ at low temperatures, being free from complicated inelastic scattering processes, reflects the electronic structure of the material. Therefore, the abrupt change in $R_H$ at 0.35K is an indication of a significant reorganization of the FS, which we believe may result from a QPT between two phases in the normal state. From the $R_H$ behavior at 0.35K we identify one phase at low x where $R_H$ changes rapidly and another phase at high x where $R_H$ varies more slowly. We find the QCP at $x_c=0.165\pm0.005$ from the intersection of best fit straight lines through the high x and low x data. This is just above the doping where $R_H$ crosses zero (somewhere between 0.16 and 0.15), a fact that can only be determined from the lowest T data. The QCP that we precisely determine from $R_H$ (and below from resistivity) is consistent with the doping trends seen in the magnetic [8], tunneling [11], and ARPES [12] measurements.

Now we identify in our resistivity measurements the QCP found in the Hall data. We fit the important low temperature range (0.35K to 20K) of the resistivity data from Figure 1 to the form $\rho(T)=\rho_0+CT^\beta$, with C, $\rho_0$ and $\beta$ independent of temperature. As an example $\rho(T)$ for x=0.17 and the fit are shown in Figure 4a. The exponent, $\beta$, obtained from the fits is presented in Figure 4b. It has a strong doping dependence and gets closer to 1 as we decrease the Ce doping from 0.19 to 0.17. Decreasing the Ce doping further to x=0.16 results in an increase in $\beta$ to 1.4. One should note that at high temperatures (T>30K) the resistivity follows a $T^2$ behavior as was previously reported [18]. At very low temperatures the temperature dependence is again $T^2$. The $T^2$ region starts below $T_0$=4.8K, 4.7K, 2K, and 6.8K ± 5% in x=0.19, 0.18, 0.17 and 0.16 respectively (for x≤0.15 the low temperature behavior is obscured by the upturn). Between the two $T^2$ regions we find the different temperature dependence with exponent 1<$\beta$<2. By fitting our data over a finite temperature range (0.35-20K) we are averaging over a Fermi liquid like $\rho\propto T^2$ region at the lowest temperatures (below $T_0$) and a quantum critical region with $\beta$<2. Thus, we obtain a fractional exponent which gets closer to 1 for x approaching $x_c$ and which goes back up towards 2 as x gets greater than $x_c$. Fournier *et al.*[18] found a linear in T resistivity from 10K down to 40mK in one of their Ce=0.17 PCCO films, a film that had $T_c$=15±4 K, somewhere between our x=0.16 and x=0.17 samples. It is possible that Fournier *et al.* hit $x_c$ in their Ce=0.17 film. Note that $x_c$ depends on both Ce and oxygen and therefore samples made by different groups can differ slightly in Ce concentration for the same carrier concentration. Also, as shown by A. Rosch [14] disorder can affect the exponent in the QCP region. Based on the value of the residual resistivity, the films of Fournier *et al.* have different disorder than ours. Taking Fournier's data into account $\beta$ appears to approach 1 around $x_c$=0.165 in our films, the same doping at which we found the kink in $R_H$. Thus, Figure 4b suggests that in a "funnel-shape" region in the doping-temperature phase diagram a linear, or close to linear, in T resistivity occurs; this funnel ends in a unique doping level, $x_c$ at T=0K. This is the behavior expected for transport properties in the quantum critical fluctuation region at finite temperatures above a QCP [2-5].

To give further support to the QPT scenario, in Figure 4c, we plot the coefficient A obtained when we fit the data to the form $\rho=\rho_0+AT^2$ in the low temperature $T^2$ region, below $T_0$, as a function of doping. From the continuity of the resistivity, A should diverge as one approaches a QCP. We find a large increase in A for x=0.17, the doping at which β has its smallest value and very close to the point where $R_H$ changes abruptly. Finally, we note that the temperature of minimum resistivity where an upturn appears in ρ(T) behaves in a similar way (Figure 4d); it decreases with increasing doping and vanishes on the overdoped side.

Our results are very similar to those found in heavy fermion materials where the resistivity has a power law dependence in temperature and this power law changes significantly with a control parameter p as p is varied across the quantum critical region [16, 23]. Along with the abrupt $R_H$ change at $x_c$=0.165 (figure 3) the power law variation in resistivity approaching 1 at the same $x_c$ is strong evidence for a QCP at $x_c$.

Note that the exponent, β=1 is expected in the marginal Fermi liquid phenomenology [13]. It is also generic that near a 2D QCP inelastic processes scale with T [2,24]. Subtle effects associated with transport present theoretical complications that at very low temperature and in very clean systems predict other powers in T [25]. However, other calculations involving systems with finite disorder [14], and the experiments in the heavy fermion systems, show that these complications are generically not significant in the accessible temperature regimes and the quantum critical predictions for β are found [16].

Very recently, Balakirev et al. [7] reported an abrupt change in Hall number near optimum doping in hole-doped $Bi_2Sr_{1-x}La_xCuO_{6+\delta}$, which they associated with a change in the FS associated with a QPT. However, no evidence from ARPES experiments were found for a such FS reorganization. In another hole-doped system, $La_{2-x}Sr_xCuO_4$, the FS changes from hole-like to electron-like at much higher doping [26].

In summary, we have presented low temperature, normal state ($H>H_{c2}$) resistivity and Hall effect data as a function of doping, which gives compelling evidence for a quantum phase transition in the electron-doped cuprate $Pr_{2-x}Ce_xCuO_{4+\delta}$. The Hall coefficient at low temperatures shows a kink near x=0.165 that suggests an abrupt change in the Fermi surface near that doping. The low temperature resistivity shows a temperature dependence with an exponent which is getting closer to 1 as one approaches $x_c$=0.165±0.05 from the overdoped or from the underdoped side. The coefficient A, found from a fit of the lowest temperature region to $\rho=\rho_0+AT^2$, increases by a factor of three at x=0.17 compared to x=0.16 and x>0.17. The upturn in the *ab*-plane resistivity vanishes around x=0.16. All of these findings strongly suggest a quantum phase transition near x=0.165. The nature of the QPT can not be determined from our transport measurements but other evidences for antiferromagnetic order suggests that in the n-doped cuprates the QPT is a magnetic one.

We thank A. J. Millis and C. M. Varma for many useful discussions and M. C. Barr for help with film growth. NSF grant DMR 01-02350 supported this work.

Fig 1.

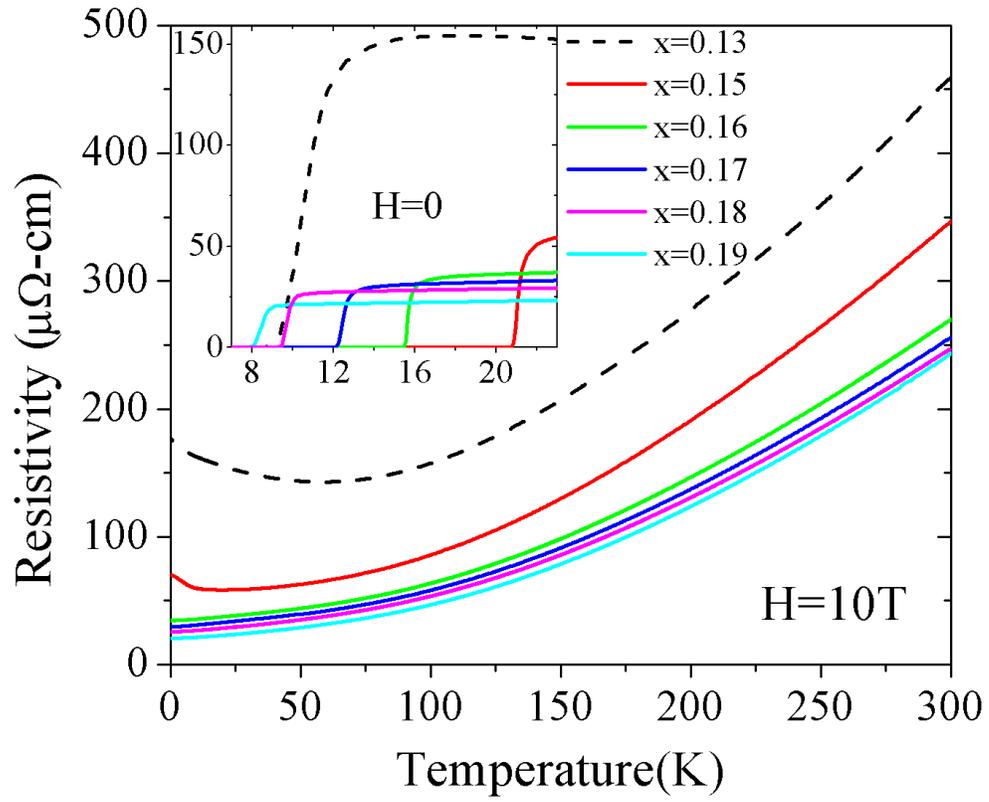

Fig. 2a

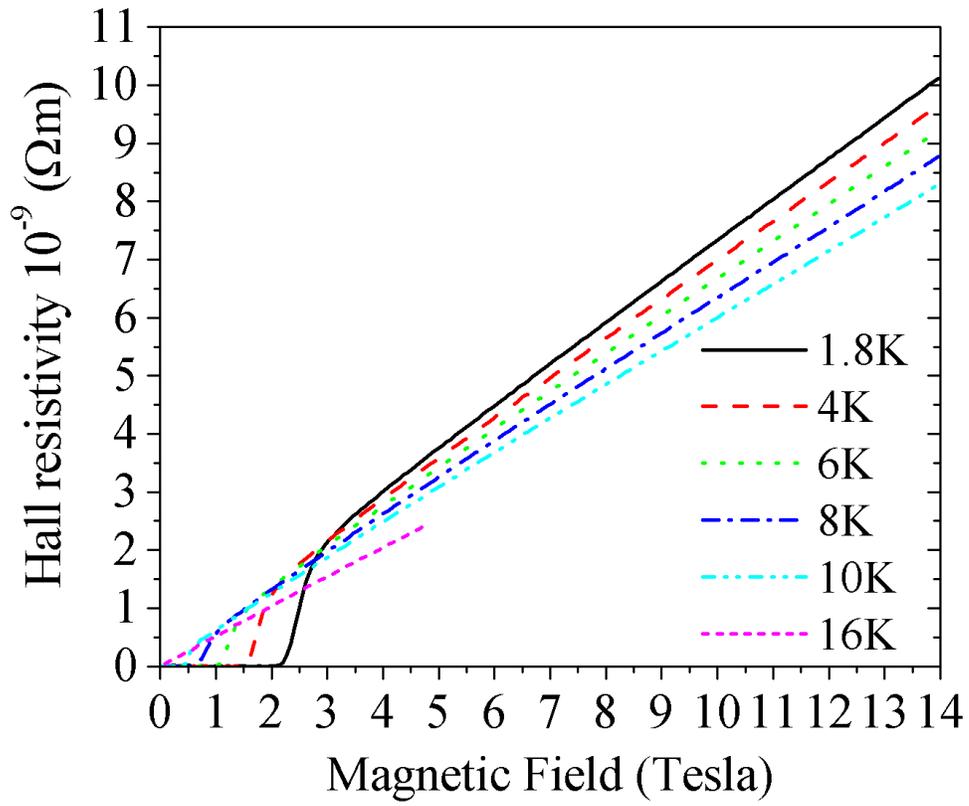

Fig. 2b

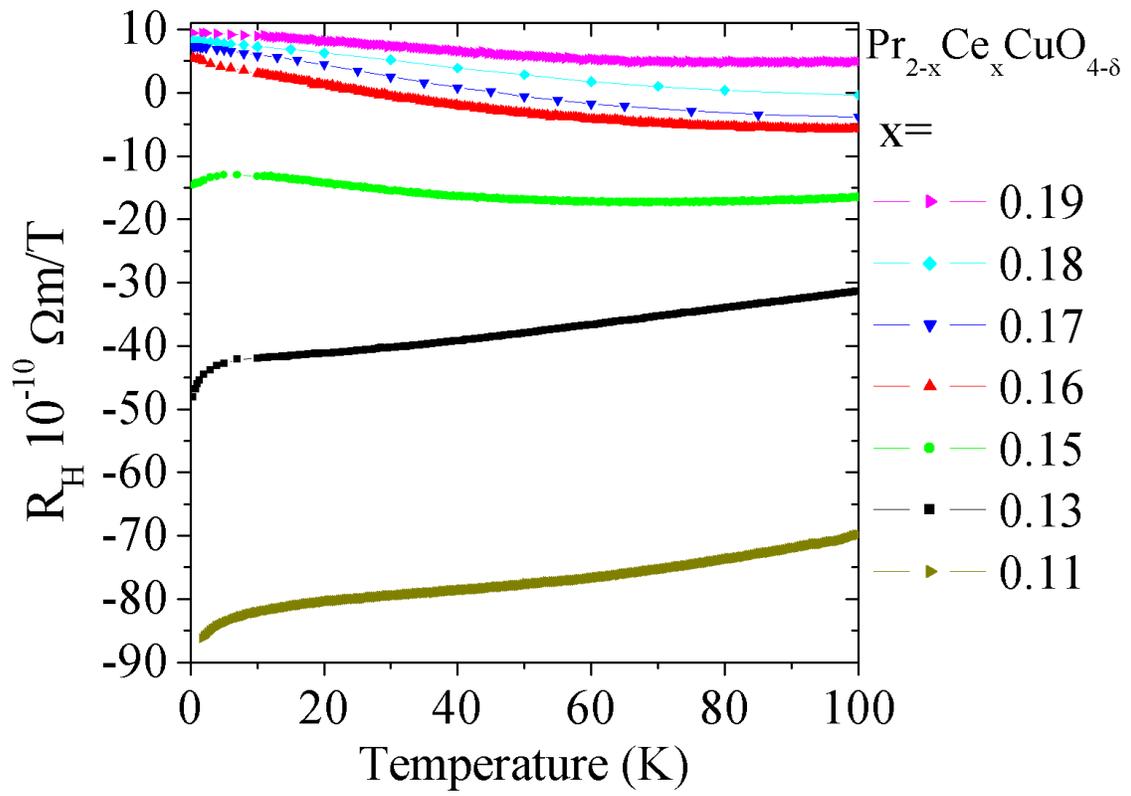

Fig. 3

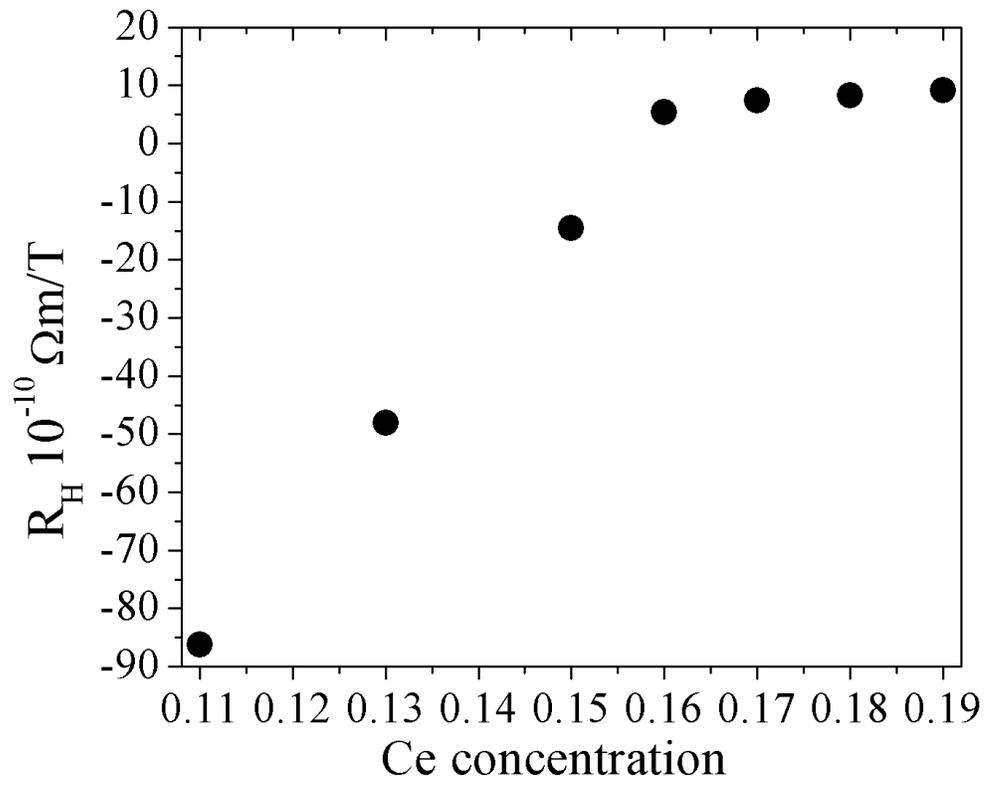

Fig. 4

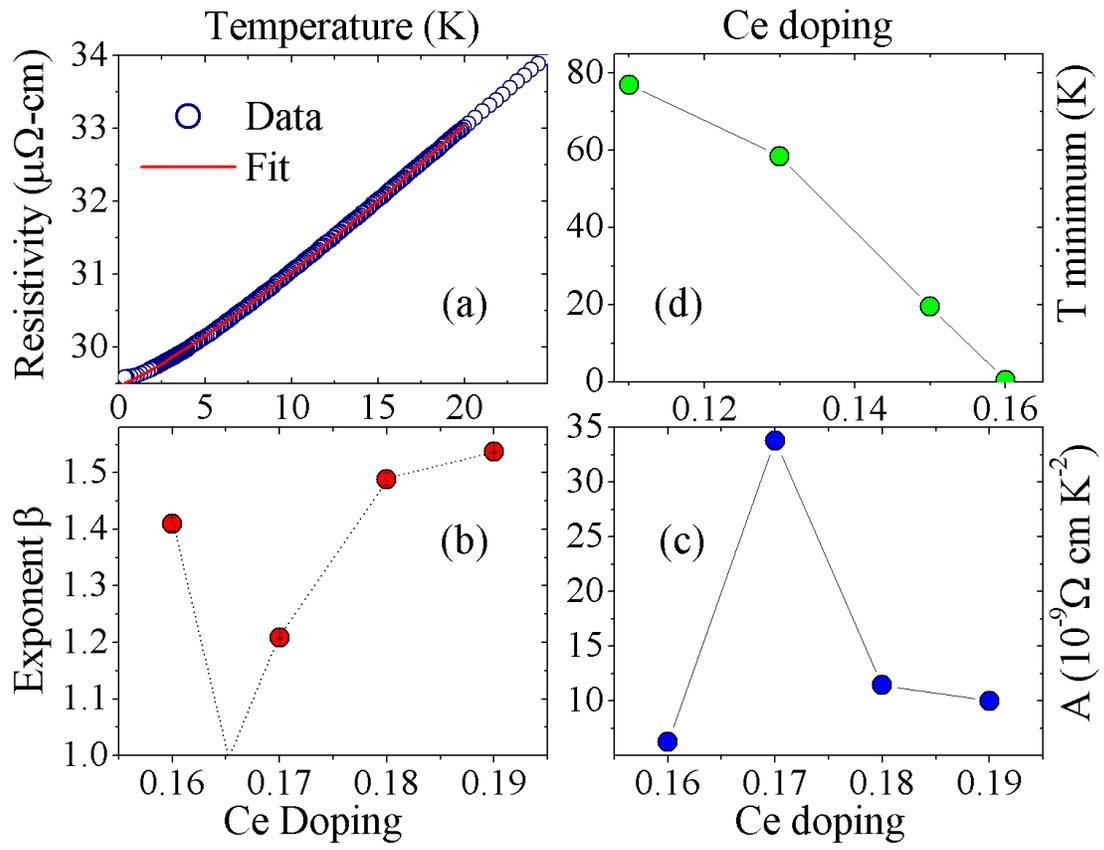

Figure Captions

Figure 1

*ab*-plane resistivity versus temperature for $Pr_{2-x}Ce_xCuO_{4+\delta}$ films with various Ce doping (x) in a magnetic field of 10T ($H>H_{c2}$) applied parallel to the c-axis. Insert: resistive superconducting transition H=0.

Figure 2

(a) Hall resistivity ($\rho_{xy}$) for the PCCO x=0.17 film. The Hall coefficient, $R_H$, is the slope of a least square fit from well above the upper critical field to 14T for $T<T_c$, and from 0 to the highest field measured, at $T>T_c$. (b) The Hall coefficient as a function of temperature for the various Ce doping.

Figure 3

The Hall coefficient at 0.35K (taken from figure 2b). A distinct kink in the Hall coefficient is seen between x=0.16 and x=0.17. The error on the concentration is approximately 0.003. The error in $R_H$ comes primarily from the error in the film thickness; it is approximately of the size of the data points.

Figure 4

(a) Analysis of the data from figure 1. We fit the data from 0.35 to 20K to $\rho=\rho_0+AT^\beta$, circles: the x=0.17 film data, solid line: fit. (b) The doping dependence of the exponent $\beta$ found from a fit from the lowest measured temperature up to 20K. The error bars are smaller than the symbols. Lines are guides to the eye. (c) The coefficient A in a fit to the form $\rho=\rho_0+AT^2$ in the low temperature $T^2$ region (described in the text) as function of doping. (d) The temperature at which minimum resistivity occurs for x≤0.15 as a function of doping. The resistive upturn is not found above 350mK for x≥0.16.

References

[1] J. Orenstein and A. J. Millis, Science **288**, 468 (2000).
[2] S. Sachdev, Rev. of Mod. Phys. **75**, 913 (2003)
[3] C. M. Varma *et al.* Phys. Rep. **361**, 267 (2002).
[4] A. Sokol and D. Pines, Phys. Rev. Lett. **71**, 2813 (1993); C. Castellani *et al*., Phys. Rev. Lett. **75**, 4650 (1995); A. Abanov and A. Chubukov, Phys. Rev. Lett., **84**, 5608 (2000).
[5] S. Chakravarty, B. I. Halperin, and D. R. Nelson, Rev. B, **39**, 2344 (1989).
[6] J. L. Tallon and J. W. Loram, Physica C, **349,** 53 (2001).
[7] F. F. Balakirev *et al.*, Nature, **424**, 912 (2003).
[8] G. M. Luke *et al.* Phys. Rev. B, **42**, 7981 (1990).
[9] H. J. Kang *et al.*, Nature, **423**, 522 (2003); M. Fujita *et al.,* cond-mat/0311269.
[10] A. Biswas *et al*., Phys. Rev. B **64**, 104519 (2001).
[11] L. Alff *et al.*, Nature **422**, 698 (2003).
[12] N. P. Armitage *et al.*, Phys. Rev. Lett. **88**, 257001 (2002).
[13] C. M. Varma *et al.*, Phys. Rev. Lett. **63**, 1996 (1989); E. Abrahams and C.M Varma, Phys. Rev B **68**, 094502 (2003)
[14] A. Rosch, Phys. Rev. B **62**, 4945 (2000).
[15] S. Chakravarty *et al.*, Phys. Rev. Lett. **89**, 277003 (2002)
[16] N. D. Mathur *et al.*, Nature, **394**, 39 (1998).
[17] G. S. Boebinger *et al.*, Phys. Rev. Lett. **77**, 5417 (1996).
[18] P. Fournier *et al.*, Phys. Rev. Lett. **81**, 4720 (1998).
[19] E. Maiser *et al.*, Physica C 297, 15 (1998).
[20] M. Naito *et al.* Physica C, **293**, 36 (1997).
[21] Wu Jiang *et al.,* Phys. Rev. Lett. 73, 1291 (1994).
[22] P. Fournier *et al.*, Phys. Rev. B **56**, 14149 (1997).
[23] J. Custers *et al.*, Nature, **424**, 524 (2003).
[24] S. Sachdev, *Quantum Phase Transition* (Cambridge University Press, Cambridge, England, 1999).
[25] R. Hlubina and T. M. Rice, Phys. Rev. B 51, 9253 (1995).
[26] A. Ino *et al*., Phys. Rev. B 65, 094504 (2002).